\documentclass[aps,prl,preprint,superscriptaddress]{revtex4-1}

\usepackage{amsmath}
\usepackage[pdftex]{graphicx}
\usepackage[thinspace]{SIunits}

\newcommand{\bb}{\mathbf b}
\newcommand{\bB}{\mathbf B}

\newcommand{\nn}{\nonumber}

\newcommand{\bs}{\mathbf{s}}

\newcommand{\bn}{\mathbf n}

\newcommand{\bS}{\mathbf{S}}
\newcommand{\be}{\begin{eqnarray}}
\newcommand{\ee}{\end{eqnarray}}
\newcommand{\la}{\langle}
\newcommand{\ra}{\rangle}

\begin{document}

\title{Three stage decoherence dynamics of electron spin qubits in an optically active quantum dot}

\author{Alexander Bechtold}
 \affiliation{Walter Schottky Institut, Technische Universit\"at M\"unchen, 85748 Garching, Germany}
\author{Dominik Rauch}
 \affiliation{Walter Schottky Institut, Technische Universit\"at M\"unchen, 85748 Garching, Germany}
\author{Fuxiang Li}
 \affiliation{Theoretical Division, Los Alamos National Laboratory, Los Alamos, New Mexico 87545, USA}
\author{Tobias Simmet}
 \affiliation{Walter Schottky Institut, Technische Universit\"at M\"unchen, 85748 Garching, Germany}
\author{Per-Lennart Ardelt}
 \affiliation{Walter Schottky Institut, Technische Universit\"at M\"unchen, 85748 Garching, Germany}
\author{Armin Regler}
 \affiliation{Walter Schottky Institut, Technische Universit\"at M\"unchen, 85748 Garching, Germany}
 \affiliation{E. L. Ginzton Laboratory, Stanford University, Stanford, California 94305, USA}
\author{Kai M\"uller}
 \affiliation{Walter Schottky Institut, Technische Universit\"at M\"unchen, 85748 Garching, Germany}
 \affiliation{E. L. Ginzton Laboratory, Stanford University, Stanford, California 94305, USA}
\author{Nikolai A. Sinitsyn}
 \affiliation{Theoretical Division, Los Alamos National Laboratory, Los Alamos, New Mexico 87545, USA}
\author{Jonathan J. Finley}
 \email{jonathan.finley@wsi.tum.de}
 \affiliation{Walter Schottky Institut, Technische Universit\"at M\"unchen, 85748 Garching, Germany}
 
\date{30 April 2015}

\begin{abstract}
\textbf{The control of discrete quantum states in solids and their use for quantum information processing is complicated by the lack of a detailed understanding of the mechanisms responsible for qubit decoherences \cite{Loss1998}. For spin qubits in semiconductor quantum dots, phenomenological models of decoherence currently recognize two basic stages \cite{Merkulov2002, Khaetskii2002, Al-Hassanieh2006}; fast ensemble dephasing due to the coherent precession of spin qubits around nearly static but randomly distributed hyperfine fields ($\sim$~2~ns) \cite{Faribault2013, Braun2005, Dou2011, Johnson2005} and a much slower process ($>$~1~$\micro$s) of irreversible relaxation of spin qubit polarization due to dynamics of the nuclear spin bath induced by complex many-body interaction effects \cite{Erlingsson2004}. We unambiguosly demonstrate that such a view on decoherence is greatly oversimplified; the relaxation of a spin qubit state is determined by three rather than two basic stages. The additional stage corresponds to the effect of coherent dephasing processes that occur in the nuclear spin bath that manifests itself by a relatively fast but incomplete non-monotonous relaxation of the central spin polarization at intermediate ($\sim$~750~ns) timescales. This observation changes our understanding of the electron spin qubit decoherence mechanisms in solid state systems.}
\end{abstract}

\maketitle

\begin{figure*}[t]
\includegraphics[width=1\textwidth]{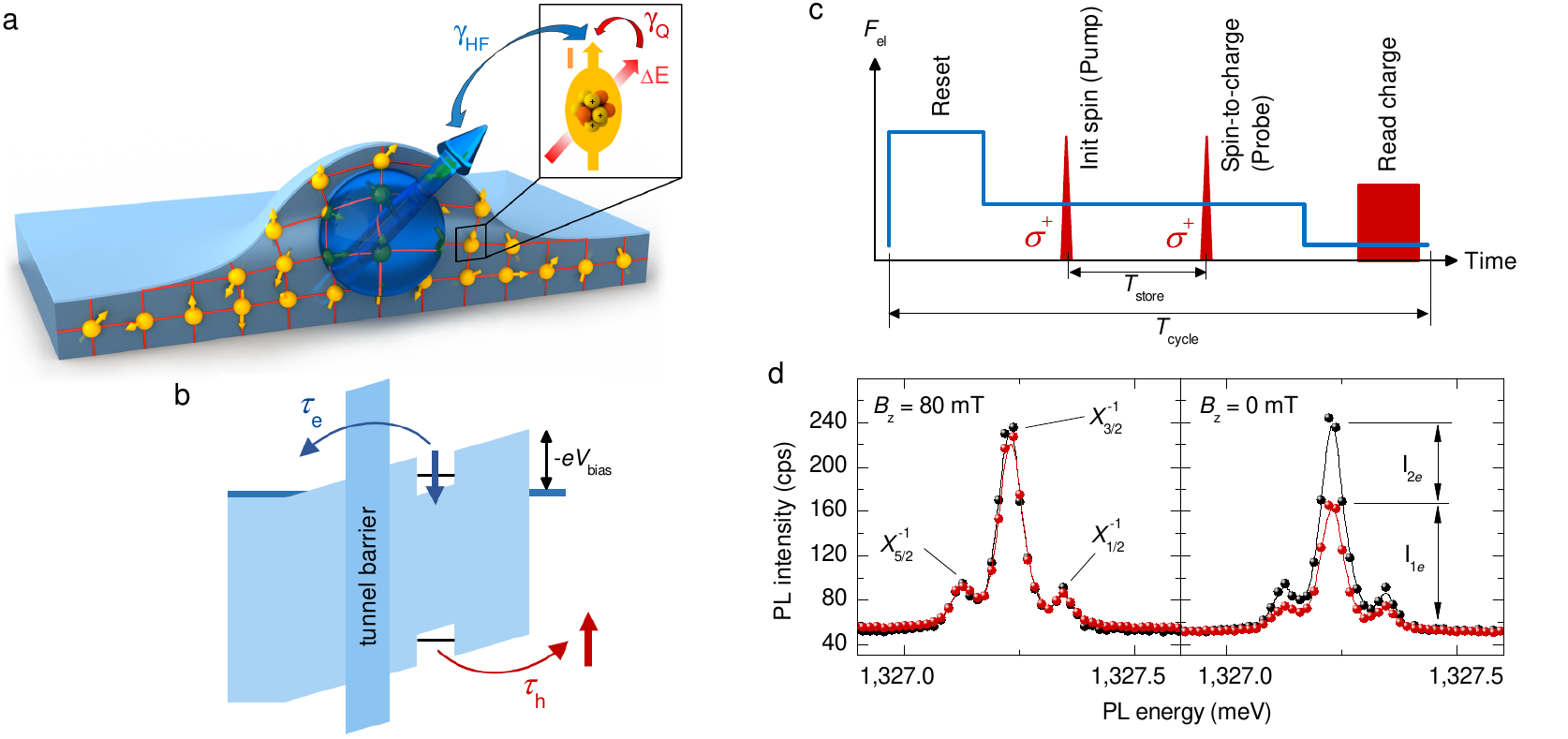}
\caption{\label{fig:Figure_1}
\textbf{Single electron spin preparation, storage and read-out.} \textbf{a,} Illustration of the hyperfine interaction ($\gamma_{HF}$) between an electron spin (blue arrow) and the nuclear spins (yellow arrows), and quadrupolar interaction ($\gamma_{Q}$) between strain induced electric field gradient $\Delta E$ and the nuclear spins $I$. \textbf{b,} Schematic representation of the band profile of the spin memory device. \textbf{c,} Representation of the applied electric field and optical pulse sequence as a function of time. The measurement cycle consists of four phases; (i) discharging the QD (Reset), (ii) electron spin preparation (Pump), (iii) spin-to-charge conversion for spin measurement (Probe) and (iv) charge read-out (Read). \textbf{d,} PL signature of the electron spin read-out for storage times of $T_\text{store}=2.8~\nano\second$. The $X_{3/2}^{-1}$ PL intensity reflects the charge state of the QD, $1e$ or $2e$, by comparison of the luminescence yield obtained with (red points) and without (black points) the application of a probe pulse. \textbf{e,} Measurement of hole ($\tau_\text{h}$) and electron tunneling time ($\tau_\text{e}$).}
\end{figure*}

A coupling of the central spin to the nuclear spin bath gives rise to an effective magnetic field, the Overhauser field \cite{Merkulov2002, Khaetskii2002, Testelin2009, Abragam1973}, in which the electron spin precesses over microsecond timescales \cite{Koppens2008, Bluhm2010, Petta2005, Press2010}. Numerous theoretical studies predicted that in the abscence of external magnetic fields the coherent character of this precession leads to a characteristic dip in the central spin relaxation, i.e. the spin polarization reaches a minimum during a few nanoseconds from which it recovers before reaching a nearly steady level at 1/3 of the initial polarization \cite{Merkulov2002, Zhang2006}. The remaining polarization has been predicted to relax slowly due to different many-body interactions in the nuclear spin bath leading to a gradual irreversible loss of coherence \cite{Khaetskii2002, Erlingsson2004, Hackmann2014, Chen2007}. Unfortunately, until now it has been impossible to test many such predictions experimentally, in particular, to observe the predicted dip in the qubit relaxation dynamics and explore phenomena occuring at much longer timescales. Here, we apply novel experimental techniques that not only clearly resolve the precession dip in the spin qubit relaxation but also provide new insights into the time-dependence of the central spin polarization during timescales that are four orders of magnitude longer than have been hitherto explored. Hereby, we utilize a spin storage device \cite{Kroutvar2004} in which a single electron spin can be optically prepared in the dot over picosecond timescales with near perfect fidelity \cite{Muller2013, Ardelt2015} (Supplementary Section 1) and stored over millisecond timescales \cite{Heiss2009}. After a well-defined storage time we \textit{directly} measure the electron spin projection along the optical axis and show that the electron spin qubit exhibits three distinct stages of relaxation. The first stage arises from a spin evolution in the randomly orientated quasi-static Overhauser field (Fig.~\ref{fig:Figure_1}a, $\gamma_\text{HF}$) inducing an inhomogeneous dephasing over the initial few nanoseconds. This is followed by an unexpected stage of the central spin relaxation, namely the appearance of a second dip in the relaxation curve after several hundred nanoseconds. We show that this unexpected feature reflects \textit{coherent} dynamic processes in the nuclear spin bath itself interpreted to arise from quadrupolar coupling (Fig.~\ref{fig:Figure_1}a, $\gamma_\text{Q}$) of nuclear spins $I$ to the strain induced electric field gradients \cite{Bulutay2012} $\Delta E$. This leads to non-ergodic fluctuations of the Overhauser field that act on the central spin. Eventually, the combined effect of quadrupolar coherent nuclear spin dynamics and incoherent co-flips of nuclear spins with the central spin induces the third stage of relaxation in which a long monotonous relaxation occurs over microsecond timescales at low magnetic fields.

The electron spin qubit studied in this work is confined in a single self-assembled InGaAs QD incorporated in the intrinsic region of a n-i-Schottky photodiode structure next to a AlGaAs tunnel barrier (Methods). As illustrated in the schematic band diagram in Fig.~\ref{fig:Figure_1}b, such an asymmetric tunnel barrier design facilitates control of the electron ($\tau_\text{e}$) and hole ($\tau_\text{h}$) tunneling time by switching the electric field inside the device. Such a control enables different modes of operation (Fig.~\ref{fig:Figure_1}c): (i) discharging the QD at high electric fields (Reset), (ii) optical electron spin initialization realized by applying a single optical picosecond polarized laser pulse (Pump), (iii) spin to charge conversion (Probe) after a spin storage time $T_\text{store}$, and (iv) charge read-out (Read) by measuring the photo luminescence (PL) yield obtained by cycling the optical $e^- \rightarrow X_{3/2}^{-1}$ transition \cite{Jovanov2011, Akimov2002, Kozin2002} (for detailed information see Supplementary Section 2). Figure~\ref{fig:Figure_1}d compares typical PL signatures of the electron spin read-out scheme for applied magnetic fields of $B_\text{z} = 80~\milli\tesla$ and $0~\milli\tesla$ for a fixed storage time of $2.8~\nano\second$. In order to measure the spin polarization $\langle S_\text{z}\rangle$ we perform two different measurement sequences: a reset-pump-read cycle (black points in Fig.~\ref{fig:Figure_1}d) to obtain the PL intensity as a reference when only one electron is present in the QD and a reset-pump-probe-read cycle (red points in Fig.~\ref{fig:Figure_1}d) from which we deduce the average charge occupation of the QD ($1e$ or $2e$) by comparing the PL intensities of the $X^{-1}_{3/2}$ ground state recombination ($I_{1e}$ or $I_{2e}$). The degree of spin polarization is then given by $\langle S_\text{z} \rangle = \left(I_{1e}-I_{2e}\right) / \left(I_{1e}+I_{2e}\right)$. As can be seen in Fig.~\ref{fig:Figure_1}d, upon reducing the magnetic field, the probability of finding the dot charged with $2e$ rises ($I_{2e}>0$) indicating that electron spin relaxation has occurred and consequently we find $\langle S_\text{z} \rangle <1$.

\begin{figure}
\includegraphics[width=0.5\textwidth]{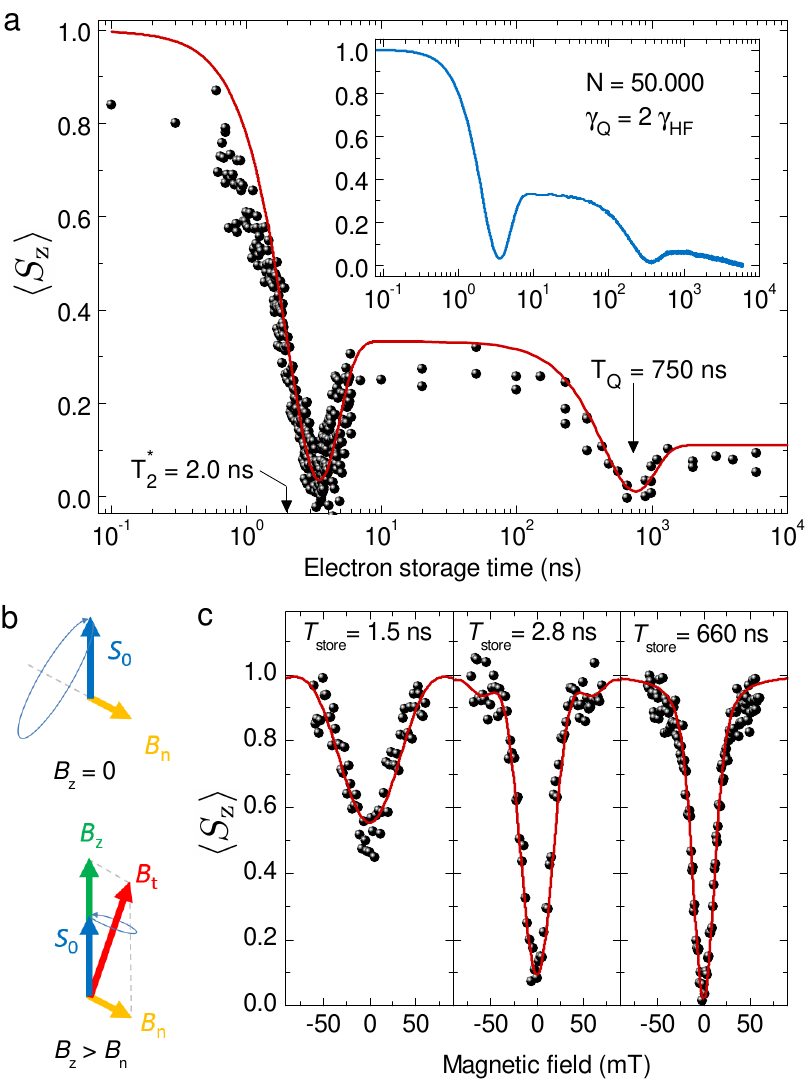}
\caption{\label{fig:Figure_3}
\textbf{Dynamics of the electron spin relaxation. a,} Experimental data reveal a fast inhomogeneous electron spin dephasing $T_2^*$ due to fluctuations of the Overhauser field while strain induced quadrupolar coupling further reduces $\langle S_\text{z}\rangle$ with a correlation time $T_{\text{Q}}$. The red line shows the result of applying a semi-classical model (Supplementary Equation S6 and S13). The inset presents a numeric calculation using a theoretical minimal-model described in the Methods and Supplementary Section 5. While the second dip is produced by quadrupolar coupling, the long relaxation tail at longer times ($>1~\micro\second$) is due to combined effects of hyperfine and quadrupolar coupling. \textbf{b,} Illustration of electron spin evolution with an initial spin $S_0$ in a total magnetic field $B_t$. \textbf{c, } Evolution of the electron spin at weak magnetic fields. The width of the dips is a result of Overhauser field fluctuations. A semi-classical model (red line) was used to simulate the experimental data (Supplementary Equation S5 and S14).} \end{figure}

The temporal evolution of $\langle S_\text{z}\rangle$ at zero external magnetic field is presented in Fig.~\ref{fig:Figure_3}a. Over the initial 20ns the average electron spin polarization exhibits a strong decay due to precession of the initial electron spin $\boldsymbol{S}_0$ around a frozen Overhauser field $\boldsymbol{B}_\text{n}$ (as schematically depicted in Fig.~\ref{fig:Figure_3}b, top). At these short timescales the Overhauser field experienced by the electron can be treated as being quasi-static but evolving between measurement cycles during the few second integration time of our experiment. The magnitude and direction of $\boldsymbol{B}_\text{n}$ are described by a Gaussian distribution function $W(\boldsymbol{B}_\text{n}) \propto \text{exp}(- \boldsymbol{B}_\text{n}^2 / 2\sigma^2_\text{n})$ with $\sigma_\text{n}$ being the dispersion of the Overhauser field \cite{Merkulov2002}. As a consequence of the field fluctuations with dispersion $\sigma_\text{n}$, the electron Larmor precession around the Overhauser field, averaged over many precession frequencies, lead to a characteristic dip in $\langle S_\text{z}\rangle$ reflecting the inhomogeneous dephasing time $T_2^*=2~\nano\second$.

In the second phase of spin relaxation observed in Fig.~\ref{fig:Figure_3}a, taking place from $20~\nano\second$ to $10~\micro\second$, the degree of spin polarization is further reduced from $\langle S_\text{z} \rangle\sim 1/3$ to a small non-vanishing value $\langle S_\text{z} \rangle\sim 1/9$. We attribute this to time dependent changes of $B_\text{n}$ due to coupling of the nuclear spins $I$ to a strain induced electric field gradient $\Delta E$ (as schematically illustrated by $\gamma_\text{Q}$ in Fig.~\ref{fig:Figure_1}a): within a nucleus the charge distribution shows deviations from a spherical symmetry, which can be described by a quadrupole moment. Due to the strain driven formation of the QDs the crystal lattice is distorted away from cubic symmetry leading to electric field gradients which couple to the quadrupolar moment of the nuclei \cite{Bulutay2012, Chekhovich2012, Chekhovich2015}. In the presence of such a quadrupolar mixing the Overhauser field acquires time dependent components which in turn modify the temporal evolution of the central electron spin \cite{Sinitsyn2012}.

In order to quantify the experimental data in Fig.~\ref{fig:Figure_3}a, we developed a semi-classical model in which the nuclear spins precess around the random static quadrupolar fields combined with a time-dependent hyperfine field of the central spin (see Supplementary). The quadrupolar coupling of a nuclear spin is characterized by the direction of the coupling axis and by the characteristic size of the energy level splittings $\gamma_\text{Q}$ along this quantization axis. In a self-assembled quantum dot, electric field gradients have a broad distribution of their direction and magnitude. We modeled them by assuming that the directions of quadrupolar coupling axes are uniformly distributed and the characteristic level splittings have Gaussian distribution throughout the spin bath: $W(\gamma_\text{Q}) \propto \text{exp}(- \gamma_\text{Q}^2 / 2\sigma^2_\text{Q})$ with $\sigma_\text{Q}$ being the single parameter that characterizes the distribution of the quadruplar coupling strengths in the spin bath.

The red line in Fig.~\ref{fig:Figure_3}a shows the prediction of this model obtained in the limit that disregards the impact of the central spin on the nuclear spin dynamics. Even in this limit the model correctly captures the appearance of both relaxation dips. While the position of the first dip is determined by the Overhauser field fluctuations ($\sigma_\text{n}=500~\micro\second^{-1}$), the second dip cannot be produced by irreversible decoherence effects. Our semi-classical model explains it as being due to coherent precession of nuclear spins around the quadrupolar axes. The Overhauser field produced by all precessing nuclear spins then fluctuates with a typical time-correlator that results in a dip at a characteristic precession frequency and saturates at a nonzero value. Throughout this regime the central spin follows the Overhauser field adiabatically, and thus its relaxation directly reflects the shape of the Overhauser field correlator. The position of the second dip is then determined only by the quadrupolar coupling strength, $T_\text{Q}=\sqrt{3}/\sigma_\text{Q}$, resulting in a value $\sigma_{Q}=2.3~\micro\second^{-1}$. Finally, in order to capture the many-body co-flip effects beyond perturbative limits within the nuclear spin bath, we performed numerical simulations of our semi-classical model including up to N$=50000$ spins. The result of these simulations at $\gamma_Q=2\gamma_{HF}$ is presented in the inset of Fig.~\ref{fig:Figure_3}a. It demonstrates that complex many-body interactions, such as spin co-flips, do not remove either of the relaxation dips provide that the quadrupolar coupling strength exceeds the hyperfine coupling $\gamma_Q>\gamma_{HF}$, as shown in Supplementary Figure S3. Co-flips, however, are responsible for the appearance of a long relaxation tail, i.e. for the irreversible third stage of central spin relaxation.

In order to obtain the Overhauser field dispersion experimentally we performed magnetic field dependent measurements of $\langle S_\text{z} \rangle$, presented in Fig.~\ref{fig:Figure_3}c. The data clearly show that $\langle S_\text{z}\rangle$ resembles a dip at low magnetic fields which can be explained as follows: in the presence of the Overhauser field $\boldsymbol{B}_\text{n}$ the electron spin precesses about the total field $\boldsymbol{B}_\text{t}= \boldsymbol{B}_\text{n}+\boldsymbol{B}_\text{z}$, as schematically depicted in Fig.~\ref{fig:Figure_3}b (bottom). At strong external magnetic fields ($B_\text{z} \gg B_\text{n}$) the total magnetic field $\boldsymbol{B}_\text{t}$ is effectively directed along $\boldsymbol{B}_\text{z}$ and the Zeeman interaction of the electron spin with the magnetic field is larger than the interaction with the Overhauser field. As a consequence, the electron spin relaxation is suppressed by an application of $B_\text{z}$ resulting in $\langle S_\text{z} \rangle \simeq 1$, as can be seen in Fig.~\ref{fig:Figure_3}c for $|B_\text{z}| > 50~\milli\tesla$. In contrast, at low external magnetic fields ($B_\text{z} \lesssim B_\text{n}$), the electron spin motion is dominated by the dynamics of the nuclear spin bath. By fitting our data using a semi-classical model (red solid line in Fig.~\ref{fig:Figure_3}c) we extract a dispersion of $\sigma_\text{n}=10.5~\milli\tesla$ that remains approximately constant at the storage times explored.

\begin{figure}
\includegraphics[width=0.5\textwidth]{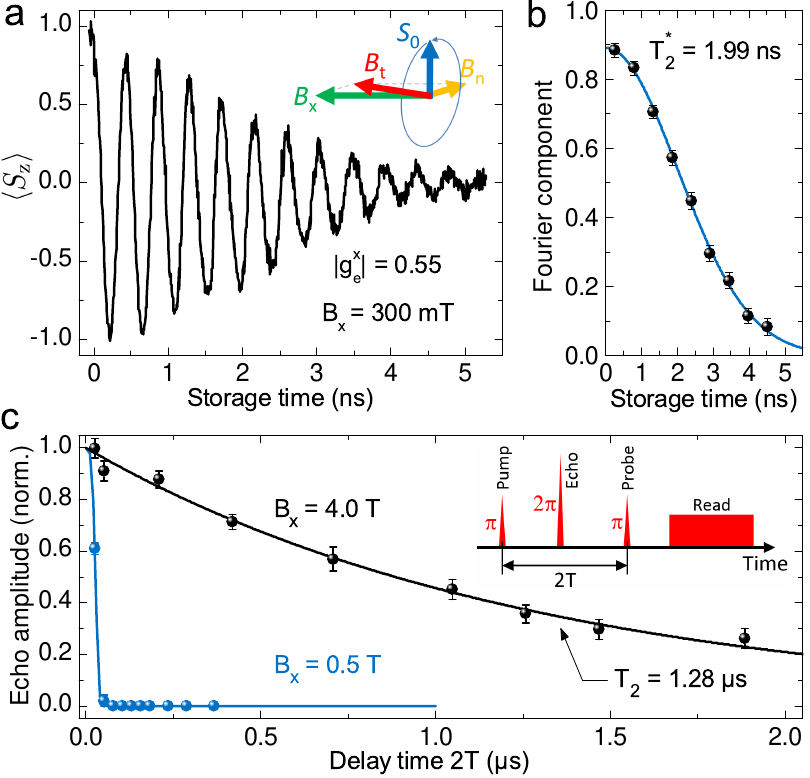}
\caption{\label{fig:Figure_2}
\textbf{Evolution of the electron spin in an in-plane magnetic field.} \textbf{a, } Free induction decay of the electron spin qubit. The Larmor oscillations are damped due to fluctuations of the Overhauser field. \textbf{b, } The Fourier component of the Larmor oscillations decays according to a Gaussian function revealing the spin dephasing time $T_2^*$. \textbf{c, } Echo signal as a function of the total evolution time at strong (black) and low magnetic fields (blue). The fit of the data are obtained using a theoretical model which is described in the Supplementary Section 4.}
\end{figure}

In addition to the out-of-plane magnetic field measurements, where the electron spin is prepared in a spin-eigenstate, we show in Fig.~\ref{fig:Figure_2}a spin-precession measurements in a fixed in-plane magnetic field. Here, the electron spin, prepared along the optical axis, precesses with the Larmor frequency ($|g_e|=0.55$) around $\boldsymbol{B}_\text{t}$, mainly directed along $\boldsymbol{B}_\text{x}$. Again, due to fluctuations of the Overhauser field, the electron spin experiences a dephasing leading to damped oscillations in the evolution of $\langle S_\text{z}\rangle$. In Fig.~\ref{fig:Figure_2}b we analyzed the Fourier component of the Larmor oscillations revealing a Gaussian envelope function indicating the Gaussian-like distribution of the Overhauser field. The variance of the fit in Fig.~\ref{fig:Figure_2}b reflects the inhomogeneous dephasing time of $1.99~\nano\second$, which is in perfect agreement with the value obtained in Fig.~\ref{fig:Figure_3}a. The dephasing time and the Overhauser field dispersion are connected via $T_2^* = \hbar / g_e \mu_\text{B} \sigma_\text{n}$. Using $g_e=0.55$ we obtain $\sigma_\text{n}=10.3~\milli\tesla$ (or $\sigma_{n}=509~\micro\second^{-1}$ in units of a relaxation rate).

In order to remove the inhomogeneous dephasing and uncover the last decoherence mechanism responsible for the long relaxation tail, we extended our pulse sequence by an additional optical (Echo) pulse which allows us the implementation of spin-echo pulse sequences. The result is shown in Fig.~\ref{fig:Figure_2}c. At strong magnetic fields ($B_\text{x}=4~\tesla$) the nuclear Zeeman splitting exceeds the quadrupolar splitting of $2.3\micro\second^{-1}$ which effectively suppresses the dephasing effect of quadrupole coupling. The dominant relaxation mechanism then stems from time dependent changes of the Overhauser field arising only from the hyperfine interaction resulting in a mono-exponential decay of the echo amplitude. From a fit we obtain a decoherence time of $T_2=1.28~\micro\second$. Upon reducing the magnetic field below the quadrupolar coupling strength ($B_\text{x}=0.5~\tesla$), the quadrupolar interaction dominates the electron spin relaxation which strongly reduces the coherence time and the echo signal follows an $\exp(-a T^4)$ behavior. The blue line in Fig.~\ref{fig:Figure_2}c shows the application of a theoretical model which is described in more detail in the Supplementary Section 4. 

In summary we investigated the electron spin decoherence mechanisms in an individual InGaAs QD over ultra-long timescales. The first relaxation phase within $T_2^* \simeq 2~\nano\second$, arising from Overhauser field fluctuations, could be suppressed using spin-echo rephasing methods. A second phase of electron spin relaxation occurs during $T_Q \simeq 750~\nano\second$ due to coherent dynamics of the nuclear spin ensemble interacting via a time-independent but non-uniform gradient of electric fields throughout the quantum dot. The third stage corresponds to irreversible processes induced by a combined effect of quadrupolar and hyperfine interaction. The good agreement between the theoretical modeling and the experimental results shows that many aspects of the electron spin decoherence at low magnetic fields are now fully understood. Our findings have major implications for predictions of the coherence time evolution of electron spin qubits at finite magnetic fields. Firstly, the comparative strengths of quadrupolar and hyperfine interactions theoretically implies that the echo coherence time of the electron spin qubits will exhibit a sudden transition with increasing magnetic field once the nuclear Zeeman splitting exeeds the quadrupolar splitting (at $\simeq 1.5~\tesla$). Secondly, while the inhomogeneous dephasing time $T_2^*$ can be reversed by the application of spin-echo methods, the strong back-action of quadrupolar couplings in the spin bath on the electron spin qubit state demonstrate the necessity of the development of \textit{strain engineered} QD structures for extended coherence times $T_2$ which may also help pave the way towards creating useful solid state quantum information devices.

\section*{Methods}
\subsection{Sample.}
The sample studied consists of a low density ($<5~\micro\meter^{-2}$) layer of nominally In$_{0.5}$Ga$_{0.5}$As-GaAs self-assembled QDs incorporated into the $d=140~\nano\meter$ thick intrinsic region of a n-i-Schottky photodiode structure. A opaque gold contact with micrometer sized apertures was fabricated on top of the device to optically isolate single dots. An asymmetric Al$_{0.3}$Ga$_{0.7}$As tunnel barrier with a thickness of $20~\nano\meter$ was grown immediately below the QDs, preventing electron tunneling after exciton generation.

\subsection{Theoretical model and numerical approximations.} 
The  minimal Hamiltonian of the central spin interacting with a nuclear spin bath with quadrupole coupling is given by
\begin{eqnarray}
\label{Ham}
 \hat{H}=  \sum_{i=1}^N \left( \gamma_{H}^i {\bf \hat{I}}_{i} \cdot {\bf \hat{S}}+g_e\mu_e  {\bf B_{\rm ex}} \cdot {\bf  \hat{S}}+ g_n\mu_n  {\bf B_{\rm ex}} \cdot {\bf  \hat{I}}_{i}+  \frac{\gamma_q^i}{2} (\hat{\bf I}_i \cdot {\bf n}_i)^2 \right),
\end{eqnarray}
where $ {\bf \hat{S}}$ is the central spin operator, $\gamma_{H}^i$ and $\gamma_q^i$ are the strengths of, respectively, the hyperfine and the quadrupole couplings of $i$th nuclear spin, $I_i>1/2$ are the sizes of the nuclear spins e.g., $I=3/2$ for Ga and $I=9/2$ for most abundant In isotopes; ${\bf n}_i$ is the unit vector along the direction of the quadrupole coupling anisotropy, which generally has a broad distribution inside a self-assembled quantum dot. 
The analytical and even numerical treatment of evolution with Eq.~(\ref{Ham}) would be too complex to achieve for a realistic number of nuclear spins $N\sim 10^5$. In order to obtain analytical and numerical estimates, we used an observation made in Ref.~\cite{Sinitsyn2012} that essential effects of the quadrupole coupling in (\ref{Ham}) are captured by a much simpler model of a spin bath with spins-1/2 only: 
\begin{equation}
\hat{H} = {\bf B} \cdot \hat{ \bf S}+\sum_{i=1}^N \gamma_H^i \hat{\bf S }\cdot \hat{\bf s}^i+\gamma_Q^i (\hat{\bf s}^i\cdot \bf n^i)+{\bf b} \cdot \hat{\bf s}^i,
\label{eff-Ham}
\end{equation}
with ${\bf B}=g_e\mu_e {\bf B}_{ex}$ and ${\bf b}=g_N\mu_N {\bf B}_{ex}$ the effective Zeeman fields acting  on, respectively, electron and nuclear spins. The spin-1/2 operators $\hat{\bf S}$  and  $\hat{\bf s}^i$ stand for the central spin and for the $i$th nuclear spin, respectively. The quadrupole coupling is mimicked here by introducing random static magnetic fields  acting on nuclear spins with the same distribution of ${\bf n}^i$ as in (\ref{Ham}), i.e, the vector ${\bf n}^i$ points in a random direction, different for each nuclear spin. Parameters $\gamma_Q^i $  are connected to $\gamma_q^i$ as $\gamma_Q^i \sim \gamma_q^i I/2$, i.e. it is the characteristic nearest energy level splitting of  a nuclear spin due to the quadrupole coupling. 

\section*{Acknowledgements}
We are very grateful to L. Cywinski for most useful and enlightening discussions. Furthermore, we gratefully acknowledge financial support from the DFG via SFB-631, Nanosystems Initiative Munich, the EU via S3 Nano and BaCaTeC. K.M. acknowledges financial support from the Alexander von Humboldt foundation and the ARO (grant W911NF-13-1-0309). Work at LANL was supported by the U.S. Department of Energy, Contract No. DE-AC52-06NA25396, and the LDRD program at LANL.

\newpage
\section*{Supplementary Information: Three stage decoherence dynamics of electron spin qubits in an optically active quantum dot}

\subsection*{1 Optical characterization of the spin memory device}

The InGaAs self-assembled quantum dots (QD) are incorporated in a spin memory device with an AlGaAs tunnel barrier immediately next to the quantum dot layer, as explained in the main article \cite{Main}. The corresponding photoluminescence (PL) spectra as a function of the electric field applied across the QD layer are presented in Fig.~\ref{fig:S1}. Here, the excitation laser is tuned to $1409~\milli\electronvolt$ exciting the charge carriers from the crystal ground state into the wetting layer from where they relax into the QD and recombine optically by producing a luminescence signature. To prevent an electron accumulation in the quantum dot due to the tunnel barrier, the sample is periodically emptied at a repetition rate of $500~\kilo\hertz$ using a reset voltage pulse (see main text\cite{Main}) of $F_\text{reset}=190~\kilo\volt/\centi\meter$ applied for $500~\nano\second$.
Directly after the reset operation, the excitation laser is applied for $1~\micro\second$ at a constant electric field which is varied from $0~\kilo\volt / \centi\meter$ to $54~\kilo\volt / \centi\meter$ to obtain the photoluminescence spectra as presented in Fig.~\ref{fig:S1}. Using this excitation sequence we observe luminescence signal from the neutral exciton $X^0$, the negatively charged exciton $X^{-1}$ and triplet $3/2$-state of the charged exciton $X^{-1}_{3/2}$ recombination for $F<30~\kilo\volt / \centi\meter$. At electric fields $F>30~\kilo\volt / \centi\meter$ the hole tunneling rate is higher than the optical recombination rate and no photoluminescence signal is produced. Thus, the photoluminescence signal in Fig.~\ref{fig:S1} is quenched for large electric fields.

\subsection*{2 Electron spin storage and readout scheme}
\begin{figure}
\includegraphics[width=1\columnwidth]{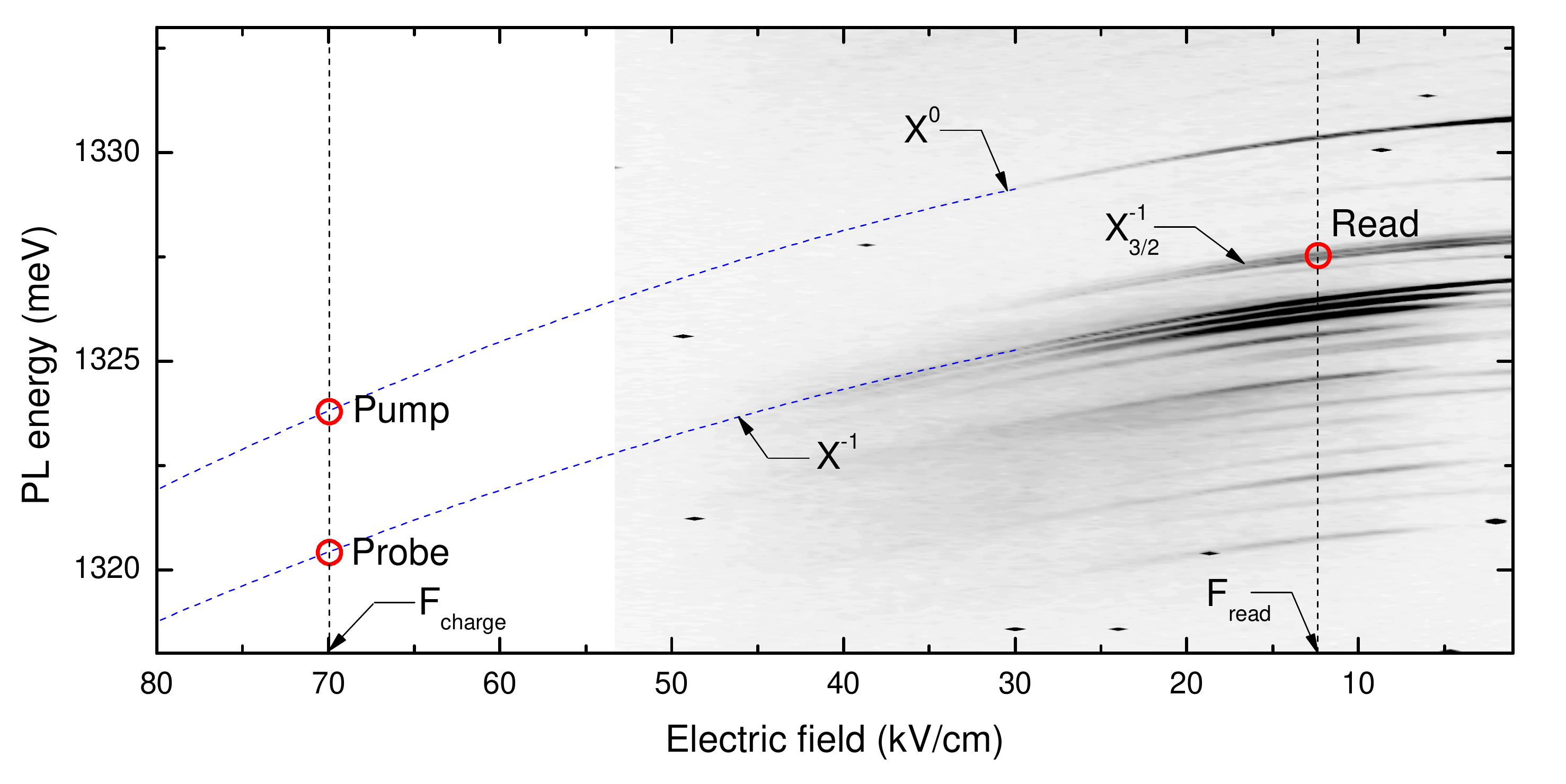}
\caption{\label{fig:S1}
Photoluminescence spectra as a function of electric field with optical excitation in the wetting layer.}
\end{figure}

In order to use the device as a spin memory we apply a time dependent electric field profile and optical pulse sequence. Initially the QD is emptied by application of high electric fields ($F_\text{reset}=190~\kilo\volt\centi\meter^{-1}$) for $500~\nano\second$. During the charging mode ($F_\text{charge}= 70~\kilo\volt\centi\meter^{-1}$) a $5~\pico\second$ duration $\sigma^+$-polarized laser pulse resonantly drives the $cgs \rightarrow X^0$ transition with $1323.8~\milli\electronvolt$ laser energy (indicated with Pump in Fig.~\ref{fig:S1}), whereupon an exciton is generated and the hole tunnels out of the QD within $\tau_h = 4~\pico\second$. The electric field is chosen such that the hole escape time is much faster than the timescale for exciton fine structure precession ($\sim150~\pico\second$) providing a spin-$\left| \downarrow \right\rangle$ initialization fidelity $\ge 98\%$. In contrast to the short hole lifetime, electron tunneling is strongly suppressed by the AlGaAs barrier leading to $\tau_\text{e} \gg 10~\micro\second$. To convert the spin information of the resident electron into a charge occupancy, a second laser pulse with $5~\pico\second$ duration and a laser energy of $1320.4~\milli\electronvolt$ is applied to resonantly excite the $1e \rightarrow X^{-1}$ transition at $F_\text{charge}=70~\kilo\volt/\centi\meter$ (indicated with Probe in Fig.~\ref{fig:S1}). During the application of this second pulse, the spin information of the resident electron is mapped into a charge occupancy of the dot. Depending on the spin projection of the initialized electron after $T_\text{store}$, the Pauli spin blockade either allows or inhibits light absorption of the second laser pulse. Thus, for electron spin-$\left| \downarrow \right\rangle$ projection the QD is charged with $1e$, whereas for spin-$\left| \uparrow \right\rangle$ the Pauli spin blockade is lifted, $X^{-1}$ creation is possible and rapid hole tunneling leaves the QD charged with $2e$. Finally, the device is biased into the charge read-out mode ($F_\text{reset}=13~\kilo\volt\centi\meter^{-1}$), where a $1~\micro\second$ duration laser pulse with a laser energy of $1350.6~\milli\electronvolt$ resonantly drives an excited state of the hot trion transition $1e \rightarrow X_{3/2}^{-1}$, probing the charge occupancy of the QD and, therefore, the electron spin polarization after $T_\text{store}$ by measuring the photo luminescence yield from the $X_{3/2}^{-1}$ recombination (indicated with Read in Fig.\ref{fig:S1}). Besides the $X_{3/2}^{-1}$ PL as shown in Fig.~1d in the main article, recombination from $X_{1/2}^{-1}$ and $X_{5/2}^{-1}$ are visible in the spectra with a reduced PL intensity. Here, the triplet states of the trion are split by the isotropic electron-hole exchange interaction ($\Delta_0=110~\micro\electronvolt$) leading to two optically bright transitions ($X_{1/2}^{-1}$, $X_{3/2}^{-1}$) and one optically dark transition ($X_{5/2}^{-1}$). The $X_{5/2}^{-1}$ transition has a non-vanishing optical recombination probability due to in-plane components of the hyperfine field that mix the spin eigenstates and is, therefore, visible in the observed PL spectra.

\begin{figure}
\includegraphics[width=0.6\columnwidth]{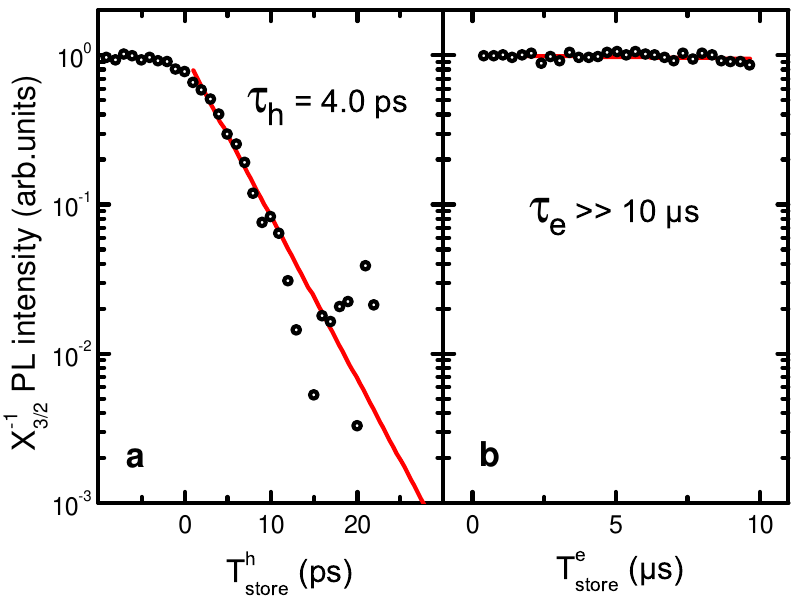}
\caption{\label{fig:S2}
Hole and Electron tunneling time. \textbf{(a)} Exciton lifetime measurement renders the hole tunneling time using a reset-pump-probe-read measurement cycle. \textbf{(b)} Electron tunneling time measurement using a reset-pump-read measurement cycle.}
\end{figure}
As explained in the main article \cite{Main}, we compare the luminescence yield obtained from the $X^{-1}_{3/2}$ optical ground state recombination using a reset-pump-read and a reset-pump-probe-read sequence to obtain the degree of electron spin polarization $\langle S_\text{z} (t)\rangle$. To obtain a high degree of the initial electron spin polarization during the charging sequence (generation of $X^0$), the hole has to be removed much faster than the exciton fine structure precession time of $\tau_\text{FSS}=150~\pico\second$. Therefore, the electric field of $F_\text{charge}=70~\kilo\volt/\centi\meter$ was chosen such that the hole tunnels out the quantum dot within $\tau_h=4~\pico\second$, while the electron remains in the quantum dot over microsecond timescales ($\tau_e \gg 10~\micro\second$). For the tunneling time measurements, shown in Fig.~\ref{fig:S2}, we use a reset-pump-probe-read sequence where the pump-laser and the probe-laser are in cross-circular configuration to prevent Pauli spin blockade (in contrast to co-circular configuration for the spin-to-charge conversion). With this sequence we monitor the $X^0$ life time by varying the time delay between the pump pulse and the probe pulse ($T^h_\text{store}$). As long as the $X^0$ is present in the QD, the probe laser is not resonant to the $e \rightarrow X^{-1}$ transition resulting in a read laser absorption whereupon a PL signal form the hot trion groundstate recombination is produced at short time delays, as can be seen in Fig.~\ref{fig:S2}a. Once the hole tunnels out the QD leaving behind a single electron, the probe laser resonantly drives the $e \rightarrow X^{-1}$ transition whereupon the read laser is off-resonant to the hot trion transition ($e \not\rightarrow X^{-1,*}_{3/2}$) and no PL signal is produced at long time delays.

The $X^0$ life time directly reflects the tunneling time of the hole ($\tau_h=4~\pico\second$) since the electron tunneling rate is negligible due to the AlGaAs tunnel barrier $\left( \frac{1}{\tau_{X^0}}=\frac{1}{\tau_{e}}+\frac{1}{\tau_{h}} \simeq \frac{1}{\tau_h}\right)$. To show that the electron life time at $F_\text{charge}=70~\kilo\volt/\centi\meter$ is sufficiently long we perform a reset-pump-read cycle and vary the time delay between pump laser and read laser ($T^e_\text{store}$). Immediately after exciton generation (pump sequence) the hole tunnels out the QD preparing the single electron. As long as the electron is present in the QD, the read laser resonantly drives the $e \rightarrow X^{-1,*}_{3/2}$ transition and PL signal from the hot trion ground state recombination can be monitored, otherwise no absorption from the read laser is possible. After a storage time of $10~\micro\second$ no electron escape is measurable, as shown in Fig.~\ref{fig:S2}b, enabling electron spin polarization measurements up to the microsecond time scale.

\subsection*{3 Theory of central spin dephasing in external out-of-plane magnetic field: short time scales}

At short time scales, $t\ll 1\mu$s, one can disregard the dynamics of nuclear spins so that the central spin polarization dynamics follow the law:
\be
\bS(t)=(\bS_0\cdot \bn)\bn+\Big(\bS_0-(\bS_0\cdot \bn)\bn \Big)\cos(\omega t) +\bS_0\times\bn \sin(\omega t), 
\label{eq:bS}
\ee
where $\bn$ is the unit vector along the total, external plus Overhauser, field:
\be
&&\bn=\frac{\bB+\bB_{n}}{\omega}, \\
&&\omega=\sqrt{(B+B_{nz})^2+B_{nx}^2+B_{ny}^2},
\ee
where we assume that the external field $\bB$ is along the measurement $z$-axis with magnitude $B$. 

The  Overhauser field  statistics is described by the Gaussian distribution \cite{Merkulov02}
 \be
W(\bB_n) \propto e^{-\frac{\bB_n^2}{2\sigma_n^2}}, \quad \sigma_n=\sqrt{N} \gamma_H,
 \label{eq:Wb}
\ee
where $N \gg 1$ is the number of  spins in the nuclear spin bath, and $\gamma_H$ is the characteristic strength of the hyperfine coupling  between a single nuclear and the central spin. 
Thus, the parameter $\sigma_n$ characterizes the typical size of the Overhauser field.

\begin{figure}\centering
\scalebox{0.6}[0.6]
{\includegraphics{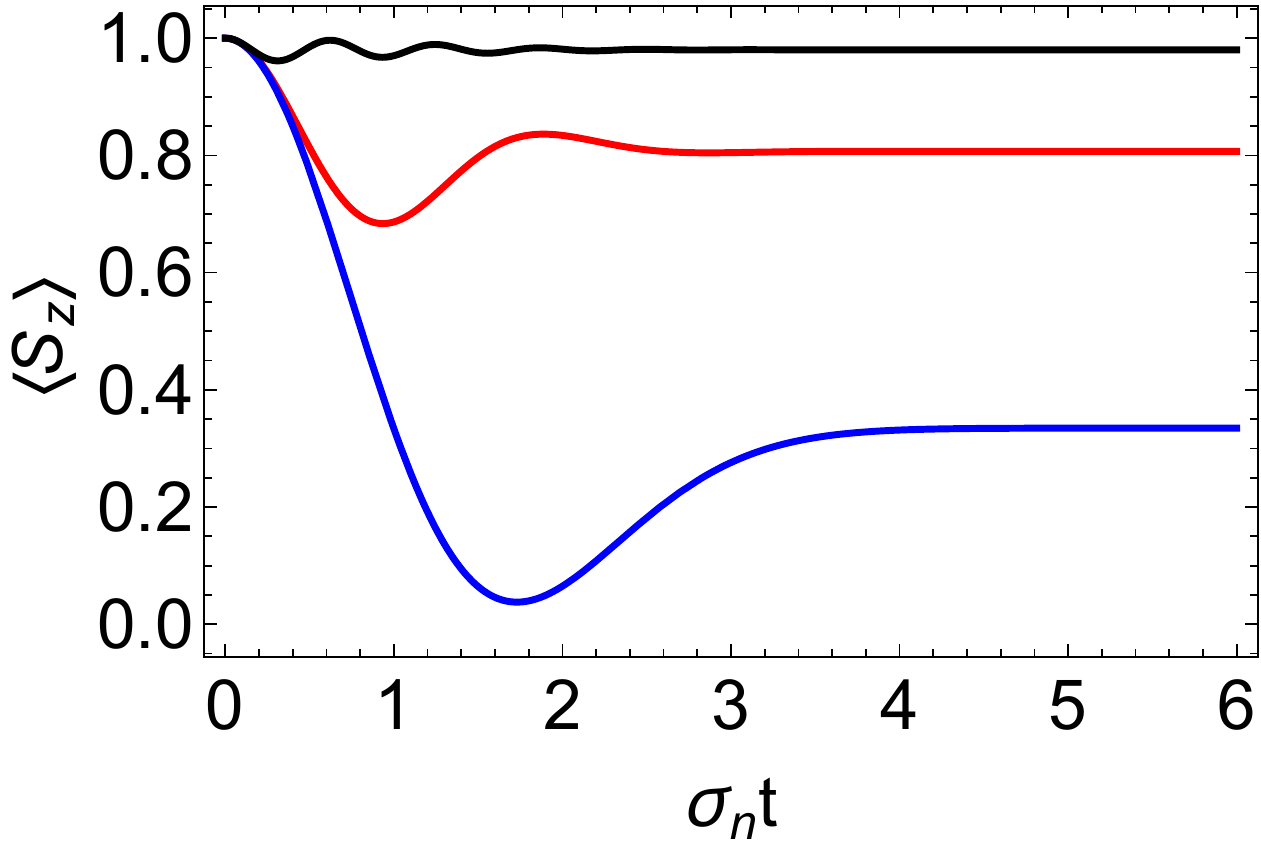}}
\caption{Time dependence of the spin polarization, $\la S_z(t) \ra$, at different external out-of-plane magnetic field values: $B=0$, $2\sigma_n$, $10 \sigma_n$ for the blue, red and black curves, respectively. 
}
\label{Szt}
\end{figure}

Averaging (\ref{eq:bS}) over (\ref{eq:Wb}) we find
\be
\la \bS(t) \ra &=& \bS_0  g(t), \quad g(t)=1-a^2+a^3 D_F\Big(\frac{1}{a} \Big) +a^2 \cos (t'/a) e^{-t'^2}  \nn \\
&&~~ -\frac{\sqrt{\pi }a^3}{4} e^{-1/a^2}\Big( {{\rm Erfi}}(1/a-it')+ {{\rm Erfi}}(1/a+it')\Big). \label{eq:g}
\ee
Here, we defined two dimensionless parameters: $a\equiv\frac{\sqrt{2}\sigma_n}{B}$ and $t'\equiv\frac{t\sigma_n}{\sqrt{2}}$.  $D_F(x)$ is the Dawson Function defined as $D_F(x)=e^{-x^2}\int_0^x e^{t^2} dt$, and ${{\rm Erfi}}(x)$ is the  imaginary error function with ${\rm Erfi(x)} = \frac{2}{\sqrt{\pi}} e^{x^2} D_F(x)$. At zero external field, $B=0$, Eq.~\ref{eq:g} reduces to the familiar formula from \cite{Merkulov02}: 
\be
\la \bS(t)\ra =\frac{\bS_0}{3}\Big(1+2(1-\sigma_n^2 t^2) e^{-\sigma_n^2t^2/2}\Big).
\ee
We used Eq.~(\ref{eq:g}) to fit experimental results (red curve) along the first dip in Fig.~2a of the main text. In addition, in Fig.~\ref{Szt} of this supplementary file, we plot $S_z(t)$ as a function of $t$ for different external magnetic field, $B=0, 2\sigma_n, 10 \sigma_n$. The spin polarization curve $\la S_z(t) \ra$ has a dip-minimum which develops at $T=\frac{\sqrt{3}}{\sigma_n}$. A comparison with the experiment gives $T=3.4$ns and $\sigma_n=\frac{\sqrt{3}}{T_{2}^*}=508\mu s^{-1}$.  

\subsection*{4 Effect of Quadrupole Coupling}
The effects of the quadrupole coupling become apparent at relatively long times $t\sim 10^2-10^3$ns. They are modeled by the Hamiltonian \cite{sinitsyn-12prl}
\begin{eqnarray}
\label{Ham}
 \hat{H}=  \sum_{i=1}^N \left( \gamma_{H}^i {\bf \hat{I}}_{i} \cdot {\bf \hat{S}}+g_e\mu_e  {\bf B_{\rm ex}} \cdot {\bf  \hat{S}}+ g_n\mu_n  {\bf B_{\rm ex}} \cdot {\bf  \hat{I}^i}+  \frac{\gamma_q^i}{2} (\hat{\bf I}_i \cdot {\bf n}_i)^2 \right),
\end{eqnarray}
where $ {\bf \hat{S}}$ is the central spin operator, $\gamma_q^i$ is the strength of the quadrupole coupling of $i$th nuclear spin, $I_i>1/2$ are the sizes of the nuclear spins, e.g., $I=3/2$ for Ga and $I=9/2$ for In most abundant isotopes; ${\bm n}_i$ is the unit vector along the direction of the quadrupole coupling anisotropy, which generally has a broad distribution inside a self-assembled quantum dot. We will make an assumption that this distribution is uniform, and test predictions of this approximation by direct comparisons with  experimental results. For simplicity, we assume a bath of spins of the same isotope, and hence identical nuclear g-factors $g_n$. 

A similar Hamiltonian has been considered previously in \cite{sinitsyn-12prl} for application to spin relaxation in hole-doped quantum dots. The major difference of electronic quantum dots is the nearly isotropic hyperfine coupling in Eq.~(\ref{Ham}), which follows from the contact exchange interaction, and a relatively strong magnitude of the hyperfine coupling, which is now comparable to the quadrupole coupling in electron-doped dots. 
This difference leads to drastically different relaxation curve for electronic spins from the merely exponential relaxation of hole spins discussed in \cite{sinitsyn-12prl}.

The analytical and even numerical treatment of evolution with Eq.~(\ref{Ham}) would be too complex to achieve for a realistic number of nuclear spins $N\sim 10^5$. In order to obtain analytical estimates for central spin dynamics, we will use an observation made in Ref.~\cite{sinitsyn-12prl} that essential effects of the quadrupole coupling in (\ref{Ham}) are captured by a much simpler model of a spin bath with spins-1/2 only: 

\be
\hat{H} = \bB \cdot \hat{ \bS}+\sum_{i=1}^N \gamma_H^i \hat{\bS }\cdot \hat{\bs}^i+\gamma_Q^i (\hat{\bs}^i\cdot \bn^i)+\bb \cdot \hat{\bs}^i,
\label{eff-Ham}
\ee
with $\bB=g_e\mu_e \bB_{ex}$ and $\bb=g_N\mu_N \bB_{ex}$ the effective Zeeman fields acting  on, respectively, electron and nuclear spins. The spin-1/2 operators $\hat{\bS}$  and  $\hat{\bs}^i$ stand for the central spin and for the $i$th nuclear spin, respectively. The quadrupole coupling is mimicked here by introducing random static magnetic fields  acting on nuclear spins with the same distribution of ${\bn^i}$ as in (\ref{Ham}), i.e, the vector $\bn^i$ points in a random direction, different for each nuclear spin. Parameters $\gamma_Q^i$ are connected to $\gamma_q^i$ as $\gamma_Q^i \sim \gamma_q^i I/2$, i.e. they characterize nearest energy level splitting of  nuclear spins due to the quadrupole coupling. 

 We are now in a position to show that, in the presence of quadrupole couplings, $\la S_z(t)\ra$ will develop a second dip with a minimum at a fraction of a microsecond.
It is easiest to see  this if we consider the case of a zero or weak external magnetic fields and a strong quadrupole coupling: $\gamma_Q \gg \gamma_H$. 
Each nuclear spin simply rotates then around the corresponding quandrupole field axis. Within the model (\ref{eff-Ham}), keeping only effects of quadrupole fields, this dynamics is given by
\be
\bs^i(t)=(\bs^i_0\cdot \bn^i)\bn^i+(\bs^i_0-(\bs^i_0\cdot \bn^i)\bn^i)\cos(\omega^i t) +\bs^i_0\times\bn^i \sin(\omega^i t),
\label{nrot}
\ee
where $\omega^i = |{\bm \gamma}_{Q}^i|$, and where we introduced the vector   ${\bm \gamma}_Q^i\equiv \gamma_Q^i \bn^i$, whose components we will choose from a Gaussian distribution:
$W({\bm \gamma}_Q) \propto e^{-\frac{{\bm \gamma}_Q^2}{2\sigma_Q^2}}$ with $\sigma_Q$ being the characteristic rms of the quadrupole coupling distribution. This choice corresponds to the uniform distribution of the anisotropy vectors $\bn^i$.
Averaging (\ref{nrot}) over the distribution of quadrupole fields, we find the correlators of the Overhauser field:
\be
 \la \bB_{n\alpha} (t)\ra =0, \quad \la \bB_{n\alpha}(t)\bB_{n\beta}(0)\ra = \gamma_H^2 \sum_i^N \la \bs^i_{\alpha}(t) \bs^i_{\beta}(0) \ra  = \sigma_n^2 \delta_{\alpha \beta } f(t),  \label{eq:bb}
\ee
with
\be
 f(t)=\frac{1}{3}\Big( 1+2(1-(\sigma_Q t)^2)e^{-\sigma_Q^2 t^2/2}\Big).
 \label{ft}
 \ee 
  At microsecond time scales, the central spin polarization follows the Overhauser field  ${\bf{B}}_n(t)$ adiabatically, i.e.,  
 \be
 \la \bS(t)\ra=\Big{\la }\bn(t)(\bn(0)\cdot\bS(0)) \Big{\ra},
 \label{scentr}
 \ee
 with $\bn(t)$ being the direction of the total field experienced by the central spin, that is, the sum of the external Zeeman coupling with the field $\bB$ and the Overhauser field $\bB_n(t)$.   
 
 For the case of $\bB=0$,  the result of averaging  (\ref{scentr}) over the initial central spin and nuclear spin states can be obtained in a closed form:
 \be
\la \bS(t)\ra = \bS(0) \Big(  \frac{2\sqrt{1-f(t)^2 }}{3\pi f(t)}  -\frac{2(1-2f(t)^2)}{3\pi f(t)^2} \sin^{-1} (f(t))\Big), \label{eq:nn}
 \ee
which shows, e.g., that $\la \bS(t) \ra$ saturates at the value $ \approx 0.095\bS(0)$ at long time and, prior to this, develops a second dip in spin relaxation at time $T_Q=\frac{\sqrt{3}}{\sigma_Q}$. This behavior is confirmed experimentally in Fig.3a of the main text. Comparison to experiment gives: $T_Q=0.75 \mu$s, which corresponds to $\sigma_Q=2.3 \mu s^{-1}$. 
  
In the presence of  an external magnetic field, the exact expression for $S_z(t)$ would be  too complex to be shown  here. Following \cite{Merkulov02}, we make the approximation that $\la \bn_{\alpha}(t) \bn_{\beta}(0) \ra \sim \frac{\la \bB_{{\rm T}\alpha}(t) \bB_{{\rm T}\alpha}(0) \ra}{\la \bB_{{\rm T} \alpha}(0) \bB_{{\rm T} \beta}(0)\ra}$ with $\bB_{\rm T}=\bB+\bB_{n}$.  We find an approximate formula: 
\be
\la S_z(t)\ra =S_z(0) \Big[ 1-a^2 f(t)+ \Big( a^3 f(t) -a(1-f(t))\Big) D_F(1/a) \Big],
\label{fit1}
\ee
where  $a$ is defined in Eq.~(\ref{eq:g}). Equation~(\ref{fit1}) produces a  good fit of the 2nd dip  in Fig.2c ($T_\text{store}=660$ns) of the main text (red curve). 

Finally, we discuss the spin echo experiment. The major reason for the decay of the spin echo, at low  values of the external magnetic field, is the noise of the Overhauser field along the applied external magnetic field. 
Let the $2\pi$ pulse be applied at time $T$ so that the echo is observed after the time $2T$. The size of the echo is then given by
\be
C_2(2T)= \Big{\la} { e}^{i \int_0^{2T} dt \alpha(t)  B_{nz} (t) } \Big{\ra},
\label{echo1}
\ee
with $\alpha(t)=1$ for $0<t<T$ and $\alpha(t)=-1$ for $T<t<2T$.
Averaging in (\ref{echo1})  is over the distribution of quadrupole fields and initial states of the nuclear spins. In the  limit of a low external magnetic field, $b\ll \sigma_c$, we find
\be
C_2(2T) = e^{-\frac{1}{4} \sigma_n^2\sigma_Q^2 T^4}.
 \label{eq:secho1}
\ee
 In the large field limit, up to the leading order terms in $\sigma_Q^2/b^2$ we find:
 \be
C_2(2T)={\rm exp} \Big[ -\frac{2\sigma_n^2  \sigma_Q^2}{b^2} \Big(  3-4e^{-\sigma_Q^2 T^2/2}  \cos (bT)  +e^{-2\sigma_Q^2 T^2} \cos(2bT) \Big) \Big]. 
\label{eq:secho2}
\ee
Equation~(\ref{eq:secho1}) shows that at low external fields, the echo signal decays non-exponentially, with a characteristic time for the relaxation of the spin echo  of the order $T_{\rm echo} \sim \sqrt{T_2^* T_Q}$, where $T_2^*$ and $T_Q$ are characteristic times of, respectively, the first dip and the second dip, defined in Fig.2a of the main text. Non-exponential relaxation of the spin echo is the direct consequence of nonergodic dynamics of nuclear spins. 
Equation~(\ref{eq:secho2}) shows that quadrupole effects on spin echo become quickly suppressed at the values of the external magnetic field exceeding the geometric mean of the Overhauser and quadrupole fields. This agrees well with the observation in Fig.~3c of the main text that relaxation of the spin echo slows down considerably at external fields $\sim 4$Tesla. Spin echo relaxes then exponentially due to the mechanisms that are essentially different from a simple nuclear spin precession. 

\subsection*{5 Details of the numerical algorithm}
An insight in a more complex regime of comparable hyperfine and quadrupole couplings can be achieved by numerical simulations of the dynamics of the central-nuclear spin density matrix within the time-dependent mean field algorithm 
developed in  \cite{sinitsyn-12prl} and \cite{dobrovitski}. Following this approach, we approximate the state vector $| \Psi \rangle$ of the total system as a product $\vert \Psi \rangle = |u_0 \rangle 
\prod_{i=1}^N |u_k \rangle$ of the
 single-spin vectors $|u_j \rangle$, ($j=0,\ldots N$). Then, at each time step we  update the state of each spin by considering its evolution with the effective Hamiltonian 
\begin{equation}
H_{\rm eff}^i = {\bm h}_i(t) \hat{{\bm \sigma}}^i,\quad i=0,1,\ldots, N,
\label{heff}
\end{equation}
where the effective field ${\bm h}_i(t)$ acting on the $i$-th spin is calculated, at each step, according to
\begin{eqnarray}
 {\bm h}_0 &=&{\bm B}+ \sum_{i=1}^N \gamma_{H}^i  \left( \sigma_z^i \hat{{\bm z}} + \sigma_x^i \hat{\bm x} +\sigma_y^i \hat{\bm y} \right), \\
 \nonumber \\
   {\bm h}_i &=& \gamma_{H}^i  \left( {\sigma}_z^0 \hat{\bm z} +\sigma_x^0 \hat{\bm x}+\sigma_y^0 \hat{\bm y} \right) + { \gamma}_Q^i  {\bm n}^i, 
\label{eq-eff}
\end{eqnarray}
where we defined
 ${\bm \sigma}^j(t)={\rm Tr} [\hat{\rho}(t) \hat{{\bm \sigma}}^j]$, and where $\hat{\rho}(t)$ is the density matrix that corresponds to the pure state $\Psi (t)$.
 
\begin{figure}
\includegraphics[width=0.5 \columnwidth]{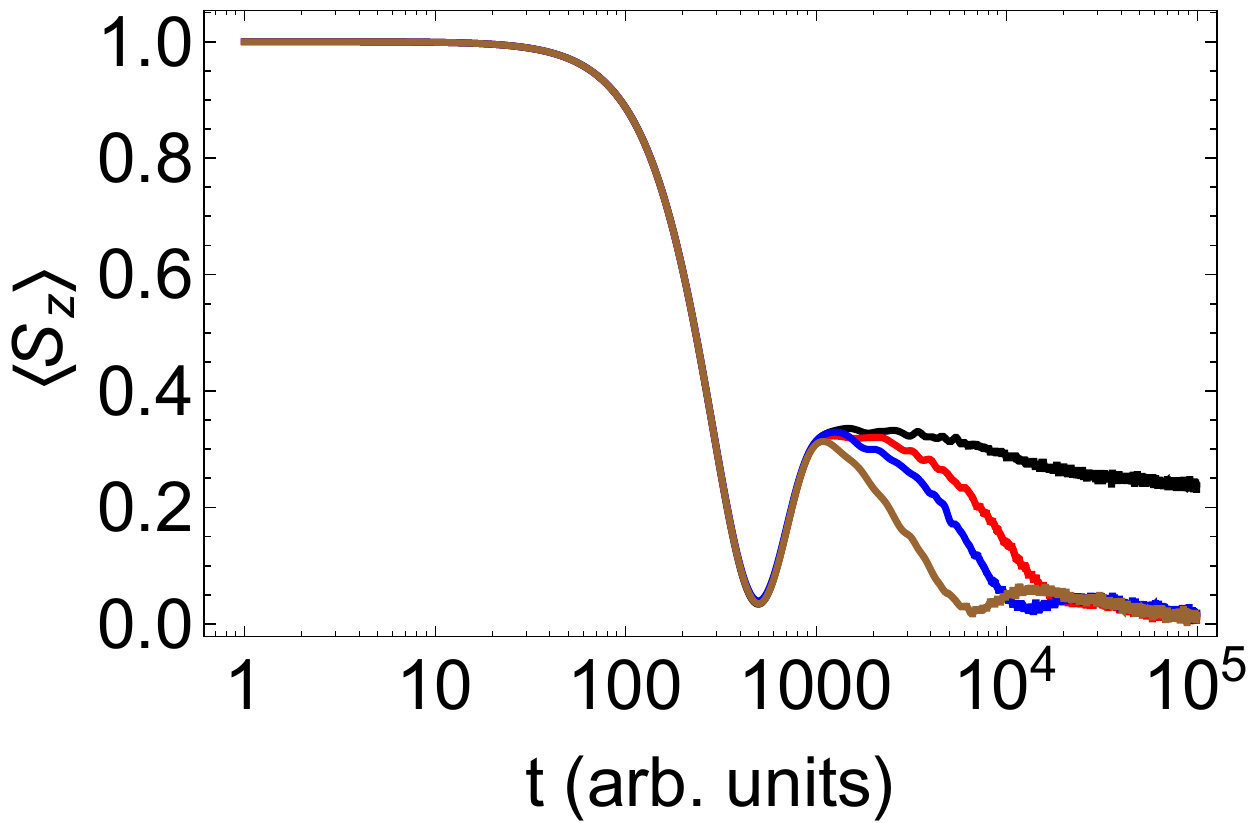}
\caption{\label{fig:sz}
Central spin relaxation at different values of quadrupole couplings $\sigma_Q$. The number of nuclear spins is $N=900$, and energy scale is set by the average value of hyperfine coupling $\gamma_H=1.0$. For individual nuclear spins, hyperfine coupling was chosen randomly from the interval ($0,2\gamma_H$).
The black, red, blue and brown curves correspond to, respectively, $\sigma_Q=0, 0.2, 0.5, 2.0$.  }
\end{figure}

The inset of Fig.~2a in the main text shows that the effect of a smaller (but comparable to quadrupole) hyperfine coupling is an enhanced relaxation at longest time scale $t \gg 1 \mu$s. At even stronger values of hyperfine couplings, the second  dip disappears. Figure~\ref{fig:sz} shows results of our numerical simulations with $N=900$ nuclear spins at $\gamma_{H}=1$ for different values of the quadrupole coupling: $\sigma_Q=0, 0.2, 0.5, 2.0$.  


\begin{thebibliography}{28}%
\bibitem{Loss1998} Loss, D. \& Divincenzo, D. P. Quantum computation with quantum dots. \emph{Phys. Rev. A} {\bf 57,} 120 (1998).
\bibitem{Merkulov2002} Merkulov, I. A., Efros, A. L. \& Rosen, M. Electron spin relaxation by nuclei in semiconductor quantum dots. \emph{Phys. Rev. B} {\bf 65,} 205309 (2002).
\bibitem{Khaetskii2002} Khaetskii, A. V., Loss, D. \& Glazman, L. Electron spin decoherence in quantum dots due to interaction with nuclei. \emph{Phys. Rev. Lett.} {\bf 88,} 186802 (2002).
\bibitem{Al-Hassanieh2006} Al-Hassanieh, K. A., Dobrovitski, V. V., Dagotto, E. \& Harmon, B. N. Numerical modeling of the central spin problem using the spin-coherent-state P representation. \emph{Phys. Rev. Lett.} \textbf{97,} 037204 (2006).
\bibitem{Faribault2013} Faribault, A. \& Schuricht, D. Spin decoherence due to a randomly fluctuating spin bath. \emph{Phys. Rev. B} {\bf 88,} 085323 (2013).
\bibitem{Braun2005} Braun, P.-F. \emph{et al.} Direct observation of the electron spin relaxation induced by nuclei in quantum dots. \emph{Phys. Rev. Lett.} {\bf 94,} 116601 (2005).
\bibitem{Dou2011} Dou, X. M., Sun, B. Q., Jiang, D. S., Ni, H. Q. \& Niu, Z. C. Electron spin relaxation in a single InAs quantum dot measured by tunable nuclear spins. \emph{Phys. Rev. B} {\bf 84,} 033302 (2011).
\bibitem{Johnson2005} Johnson, A. C. \emph{et al.} Triplet–singlet spin relaxation via nuclei in a double quantum dot. \emph{Nature} {\bf 435,} 925-928 (2005).
\bibitem{Erlingsson2004} Erlingsson, S. \& Nazarov, Y. Evolution of localized electron spin in a nuclear spin environment. \emph{Phys. Rev. B} {\bf 70,} 205327 (2004).
\bibitem{Testelin2009} Testelin, C., Bernardot, F., Eble, B. \& Chamarro, M. Hole-spin dephasing time associated with hyperfine interaction in quantum dots. \emph{Phys. Rev. B} {\bf 79,} 195440 (2009).
\bibitem{Abragam1973} Abragam, A. The principles of nuclear magnetism. (Clarendon Press, Oxford, 1973).
\bibitem{Koppens2008} Koppens, F. H. L., Nowack, K. C. \& Vandersypen, L. M. K. Spin echo of a single electron spin in a quantum dot. \emph{Phys. Rev. Lett.} {\bf 100,} 236802 (2008).
\bibitem{Bluhm2010} Bluhm, H. \emph{et al.} Dephasing time of GaAs electron-spin qubits coupled to a nuclear bath exceeding 200$\mu$s. \emph{Nature Phys.} {\bf 7,} 109-113 (2010).
\bibitem{Petta2005} Petta, J. R. \emph{et al.} Coherent manipulation of coupled electron spins in semiconductor quantum dots. \emph{Science} {\bf 309,} 2180 (2005).
\bibitem{Press2010} Press, D. \emph{et al.} Ultrafast optical spin echo in a single quantum dot. \emph{Nature Photon.} {\bf 4,} 367-370 (2010).
\bibitem{Zhang2006} Zhang, W., Dobrovitski, V. V., Al-Hassanieh, K. A., Dagotto, E. \& Harmon, B. N. Hyperfine interaction induced decoherence of electron spins in quantum dots. \emph{Phys. Rev. B} {\bf 74,} 205313 (2006).
\bibitem{Hackmann2014} Hackmann, J. \& Anders, F. B. Spin noise in the anisotropic central spin model. \emph{Phys. Rev. B} {\bf 89,} 045317 (2014).
\bibitem{Chen2007} Chen, G., Bergman, D. L. \& Balents, L. Semiclassical dynamics and long-time asymptotics of the central-spin problem in a quantum dot. \emph{Phys. Rev. B} {\bf 76,} 045312 (2007).
\bibitem{Kroutvar2004} Kroutvar, M. \emph{et al.} Optically programmable electron spin memory using semiconductor quantum dots. \emph{Nature} {\bf 432,} 81-84 (2004).
\bibitem{Muller2013} M\"{u}ller, K. \emph{et al.} All optical quantum control of a spin-quantum state and ultrafast transduction into an electric current. \emph{Sci. Rep.} {\bf 3,} 1906 (2013).
\bibitem{Ardelt2015} Ardelt, P.-L. \emph{et al.} Controlled tunneling induced dephasing of Rabi rotations for ultra-high fidelity hole spin initialization. \emph{ArXiv e-preprints} 1504.00807 (2015).
\bibitem{Heiss2009} Heiss, D. \emph{et al.} Selective optical charge generation, storage, and readout in a single self-assembled quantum dot. \emph{Appl. Phys. Lett.} {\bf 94,} 072108 (2009).
\bibitem{Bulutay2012} Bulutay, C. Quadrupolar spectra of nuclear spins in strained In$_x$Ga$_{1-x}$As quantum dots. \emph{Phys. Rev. B} {\bf 85,} 115313 (2012).
\bibitem{Jovanov2011} Jovanov, V., Kapfinger, S., Bichler, M., Abstreiter, G. \& Finley, J. J. Direct observation of metastable hot trions in an individual quantum dot. \emph{Phys. Rev. B} {\bf 84,} 235321 (2011).
\bibitem{Akimov2002} Akimov, I. A., Hundt, A., Flissikowski, T. \& Henneberger, F. Fine structure of the trion triplet state in a single self-assembled semiconductor quantum dot. \emph{Appl. Phys. Lett.} {\bf 81,} 4730 (2002).
\bibitem{Kozin2002} Kozin, I. E. \emph{et al.} Zero-field spin quantum beats in charged quantum dots. \emph{Phys. Rev. B} {\bf 65,} 241312 (2012).
\bibitem{Chekhovich2012} Chekhovich, E. A. \emph{et al.} Structural analysis of strained quantum dots using nuclear magnetic resonance. \emph{Nature Nanotech.} {\bf 7,} 646-650 (2012).
\bibitem{Chekhovich2015} Chekhovich, E. A., Hopkinson, M., Skolnick, M. S. \& Tartakovskii, A. I. Suppression of nuclear spin bath fluctuations in self-assembled quantum dots induced by inhomogeneous strain. \emph{Nat. Commun.} {\bf 6,} 6348 (2015).
\bibitem{Sinitsyn2012} Sinitsyn, N. A., Li, Y., Crooker, S. A., Saxena, A. \& Smith, D. L. Role of nuclear quadrupole coupling on decoherence and relaxation of central spins in quantum dots. \emph{Phys. Rev. Lett.} {\bf 109,} 166605 (2012).
\end{thebibliography}

\begin{thebibliography}{4}%
\bibitem{Main}   Bechtold, A. \emph{et al.} Three stage decoherence dynamics of electron spin qubits in an optically active quantum dot. \emph{Main article}.
\bibitem{Merkulov02} Merkulov, I. A., Efros, A. L. \& Rosen, M. Electron spin relaxation by nuclei in semiconductor quantum dots. \emph{Phys. Rev. B} {\bf 65,} 205309 (2002).
\bibitem{sinitsyn-12prl} Sinitsyn, N. A., Li, Y., Crooker, S. A., Saxena, A. \& Smith, D. L. Role of nuclear quadrupole coupling on decoherence and relaxation of central spins in quantum dots. \emph{Phys. Rev. Lett.} {\bf 109,} 166605 (2012).
\bibitem{dobrovitski} Al-Hassanieh, K. A., Dobrovitski, V. V., Dagotto, E. \& Harmon, B. N. Numerical modeling of the central spin problem using the spin-coherent-state P representation. \emph{Phys. Rev. Lett.} \textbf{97,} 037204 (2006).
\end{thebibliography}
\end{document}